\documentclass[12pt]{iopart}   

\usepackage{graphicx}

\begin{document}

\title{Introduction to nonlinear discrete systems: Theory and modeling}

\author{E. N. Tsoy$^1$ and B. A. Umarov$^2$}

\address{$^1$Physical-Technical Institute of the Uzbek Academy of Sciences,\\
Bodomzor yuli st. 2-B, Tashkent, 100084, Uzbekistan\\
$^2$Department of Physics, Faculty of Science,\\ International Islamic
University Malaysia, 25200 Kuantan, Malaysia }

\vspace{10pt}
\begin{indented}
\item[] 23 February 2018
\end{indented}

\begin{abstract}
   An analysis of discrete systems is important for understanding of
various physical processes, such as excitations in crystal lattices and
molecular chains, the light propagation in waveguide arrays, and the
dynamics of Bose-condensate droplets. In basic physical courses, usually
linear properties of discrete systems are studied. In this paper we propose
a pedagogical introduction to the theory of nonlinear distributed systems.
The main ideas and methods are illustrated using a universal model for
different physical applications, the discrete nonlinear Schr\"{o}dinger
(DNLS) equation. We consider solutions of the DNLS equation and analyze
their linear stability. The notions of nonlinear plane waves, modulational
instability, discrete solitons and the anti-continuum limit are introduced
and thoroughly discussed. A Mathematica program is provided for better
comprehension of results and further exploration. Also, few problems,
extending the topic of the paper, for independent solution are given.
\end{abstract}

\vspace{2pc}

\submitto{\EJP}

\section{Introduction}
\label{sec:intro}

   A discrete system is a system that consists of several (or an infinite
number of) well-separated points (sites). Each site is characterized by
some variables, so that at a given time these variables specify a state of
the system. A change of variables on each site may depend on values on
other sites.

  Many important physical systems, such as crystal lattices, atomic chains,
polymers, arrays of resonators and waveguides, electrical transmission
lines, and spin lattices are essentially
discrete~\cite{Kivs03,Ablo04,Flac08,Lede08,Kevr09,Kart11}. A recent advance
in this field is summarized in books and research papers, that are written
mainly for specialists. The purpose of this paper is to give an accessible
introduction to the subject, and to illustrate the analytical methods used.

  Different aspects of discrete systems are discussed in educational
literature (see e.g.~\cite{Daux05,Leve06,Zhan10,Newm11,Lian15}) to enhance
the corresponding university courses. However, mainly linear properties are
studied in such courses. In this work, the basic ideas of the nonlinear
dynamics are presented. We demonstrate main steps for analysis of discrete
systems, using a well-studied model, the discrete nonlinear Schr\"{o}dinger
(DNLS) equation~\cite{Kivs03,Ablo04,Flac08,Lede08,Kevr09,Kart11}. We analyze
properties of  plane waves and localized modes. We also provide a script in
Mathematica that helps to do the analysis. A student can test these methods
on the DNLS equation, and then extend and modify them to study more complex
systems. Our paper can also be useful for lecturers, who want to include this
topic to courses on nonlinear physics.

  The DNLS equation is written in the following form:
\begin{equation}
  i {d u_k \over d t} + \beta_0 u_{k} + \beta (u_{k+1}  + u_{k-1}) +
    \gamma |u_k|^2 u_k = 0,
\label{dnls}
\end{equation}
where $u_k$ is the field value at $k$-th point, $k$ is an integer number,
$\beta > 0$ characterizes a coupling with neighbor sites, and $\gamma$ is
the nonlinearity parameter. In this Section, we give examples of different
physical systems described by Eq.~(\ref{dnls}).

\begin{figure}[htb]
\centerline{
\includegraphics[width= 6.5cm]{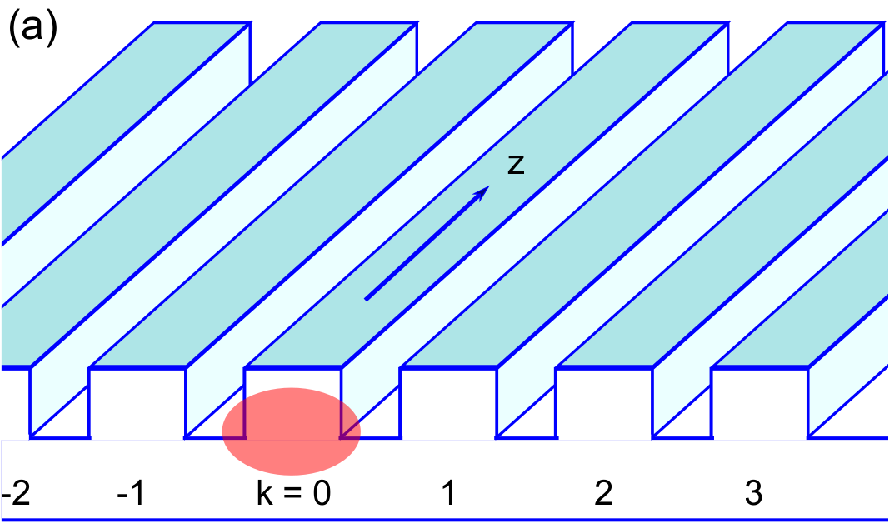} \hspace{5mm}
\includegraphics[width= 6.5cm]{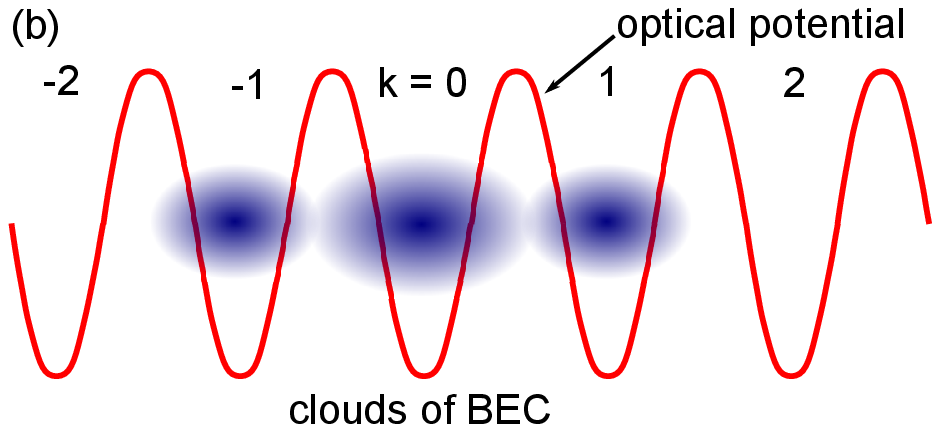}}
\vspace{0.5cm}
\caption{Examples of systems, described by the DNLS equation. (a) An array of
optical rib waveguides. Light in a waveguide (a spot at $k = 0$) is coupled
with nearest neighbor waveguides. (b) Clouds of Bose-Einstein condensate,
distributed in minima ($k = -1, 0$ and 1) of the periodic optical potential.
}
\label{fig:sys}
\end{figure}

  Let us consider an array of optical waveguides, see
Fig.~\ref{fig:sys}(a). The field variable $u_k$ corresponds to the electric
field $E$ in $k$-th waveguide. Waveguides are situated close to each other,
so that light in one waveguide can tunnel into nearest-neighbor waveguides.
This effect induces an overlap of modes and interaction of fields in the
adjacent waveguides (see the third term in Eq.~(\ref{dnls})). It is assumed
that the refractive index $n$ of the material is intensity-dependent, $ n =
n_0 + n_2 |E|^2$. Such a dependence of $n$ results in linear (the second)
term and nonlinear (the fourth) term in Eq.~(\ref{dnls}). Therefore,
Eq.~(\ref{dnls}) describes light propagation in a waveguide array. The
equation can be derived from the Maxwell equations in the limit of weakly
coupled modes~\cite{Kivs03,Lede08,Kart11}. Parameter $\beta_0$ is the
linear propagation constant, $\beta$ is the coupling coefficient, and
$\gamma \sim n_2$ characterizes the Kerr nonlinearity. The propagation
distance $z$ plays here the role of time. A study of beam dynamics in
arrays can be useful for light routing in photonic circuits, for beam
steering and switching~\cite{Kivs03,Lede08,Kart11}. Waveguide arrays can
serve as a universal testbed for analysis of phenomena from solid state
physics, such as Anderson localization and Bloch oscillations.

  Now, we consider a gas of ultra-cold (e.g. rubidium) atoms at temperature
$\sim 10^{-7}$~K in an optical lattice, see Fig.~\ref{fig:sys}(b). Because
of such a low temperature, all atoms are in the ground state, and they form
a new state of matter, a Bose-Einstein condensate (BEC)~\cite{Peth08}. An
optical lattice, which is a standing wave created due to interference of
laser beams, forms a periodic potential for a BEC. Let us analyze a BEC
distributed in the minima of the potential. A relative number of atoms in
the $k$-th minimum are described by $|u_{k}|^2$. Then, assuming that
particles on different minima interact weakly due to tunneling, and taking
into account the two-body scattering effect, one can arrive to
Eq.~(\ref{dnls}). More rigorous derivation of Eq.~(\ref{dnls}) from the
Gross-Pitaevsky equation for a BEC in a periodic potential can be found in
Ref.~\cite{Trom01}. A BEC in optical lattices, as an ensemble of coherent
objects, is actively studied from a fundamental point of view. A set of
interacting condensate droplets can be used in experiments on matter wave
interference, and precise measurements~\cite{Peth08}.

   The DNLS Eq.~(\ref{dnls}) describes also other systems, such as
Josephson junction arrays, layered magnetic systems, organic molecules and
DNA, see~\cite{Flac08,Kevr09,Kart11}. Since the DNLS equation has different
applications, in this work we treat it in general terms, referring to $u_k$
as the field variable and to $t$ as time. Now, when the physical importance
of Eq.~(\ref{dnls}) is justified, let us consider its main properties.

   Equation~(\ref{dnls}) has two conserved quantities
\begin{equation}
  P = \sum_{k=1}^{N}{|u_k|^2},\quad
  H = \sum_{k=1}^{N} \left( \beta_0 |u_k|^2 + \beta |u_k - u_{k-1}|^2 +
    {\gamma \over 2} |u_k|^4 \right).
\label{invar}
\end{equation}
In the context of waveguide arrays, $P$ is the total power, while $H$ is
the Hamiltonian. The second term in Eq.~(\ref{dnls}) can be eliminated by
using transformation $u_k(t) \to u_k(t) \exp(i \beta_0 t)$. In the rest of
the paper, we assume $\beta_0 = 0$.

  When $\gamma = 0$, an excitation, initially localized on a single site,
spreads across the array due to coupling. Nonlinearity ($\gamma \neq 0)$
can support localized states, in which the energy is locked mainly
in few sites. These localized states are called discrete solitons (DSs).
There are two basic types of solitons, namely, bright soltions and
dark solitons. The names come from applications in optics. A bright DS
corresponds to a distribution that vanishes far from the mode center.
Such a mode is observed a set of bright spots located in few wavegudes.
A dark DS corresponds to a dip on a constant background. We focus our
attention on bright DSs.

  Plane waves and DSs are considered as fundamental modes of nonlinear
discrete systems. Any initial field distribution with finite $P$ ends up
typically in a set of spreading waves and DSs~\cite{Ablo04,Kevr09} .
Therefore, in order to study the dynamics of discrete systems, one need to
understand properties of these excitations.

\section{Analysis of waves in discrete systems}
\label{sec:awds}

\subsection{Plane waves}
\label{sec:pw}

   For discrete systems, it is instructive to start the analysis from a
plane wave solution. We look for a solution in the following form:
\begin{equation}
  u_k(t) = a \exp[i (q k - \omega t)],
\label{plane}
\end{equation}
where $a$, $q$, and $\omega$ are the amplitude, the wave number and the
frequency of the plane wave. A substitution of Eq.~(\ref{plane}) into
Eq.~(\ref{dnls}) results in the following dispersion relation:
\begin{equation}
  \omega = -(2 \beta \cos q + \gamma a^2).
\label{ldisp}
\end{equation}
This is an important characteristics of plane waves. The physical meaning of
the dispersion relation is simple. If at $t = 0$, one prepares a profile in a
form of Eq.~(\ref{plane}) with given $q$, then the real and imaginary parts
oscillate in time with frequency $\omega$. Alternatively, exciting in time a
single site with frequency $\omega$, one generates a wave with wavenumber
$q$ defined by Eq.~(\ref{ldisp}).

   First, we consider the dispersion relation of the linear system, $\gamma
= 0$ (or $a \to 0$). The first derivative $d\omega / dq$ defines the group
velocity, while the second derivative $d^2 \omega / dq^2$ is the wave
dispersion. Since, in general, $d^2 \omega / dq^2 \neq 0$, we conclude that
the inter-site coupling induces effective dispersion. This dispersion is
responsible for a spreading of initially localized distribution. Another
property, worth to notice, is that dispersion can change its sign,
depending on the value of $q$. We return to this property in the discussion of
localized waves.

   In presence of nonlinearity, $\gamma \neq 0$, the dispersion relation
shifts up or down depending on sign of $\gamma$. The dependence of the wave
frequency on the amplitude is typical for nonlinear systems.

  For $q = 0$ ($q = \pi$), amplitudes $a e^{i q k }$ in nearest-neighbor
sites have the same (opposite) sign. The corresponding distribution is called
unstaggered (staggered).

  The next step of the analysis is to study the dynamics of small
perturbations of the plane wave. A growth of small perturbations is a
manifestation of modulational instability (MI). Usually, the result of MI
is a structure with well-separated localized waves that can be associated
with DSs.

  In order to find unstable wave parameters, we represent the solution in
the following form:
\begin{equation}
  u_k(t) = [a + b_k(t)] \exp[i (q k - \omega t)],
\label{pert}
\end{equation}
where $b_k(t)$ is a complex amplitude of small modulations
of the plane wave. Substituting Eq.~(\ref{pert}) into Eq.~(\ref{dnls}), and
taking only first-order terms on $b_k$, we get the following linear
equation for perturbations:
\begin{equation}
  i {d b_k \over d t} + \omega b_k + \beta (b_{k-1} e^{-i q} + b_{k+1} e^{i q})
  + \gamma(2 a ^2 b_k + a^2 b_k^{*}) = 0,
\label{pertdyn}
\end{equation}
Separating the real and imaginary parts of $b_k$, and representing them
$\sim \exp[i(Q n - \Omega t)]$, we obtain a linear algebraic system. The
compatibility condition of this system gives the dispersion relation of
perturbations, cf.~\cite{Kevr09}:
\begin{equation}
   (\Omega - 2\beta \sin Q \sin q)^2 = 8\beta \sin^2(Q/2) \cos q\,
     [2 \beta \sin^2(Q/2) \cos q - \gamma a^2].
\label{ds_pert}
\end{equation}
When the right hand side of Eq.~(\ref{ds_pert}) is negative, $\Omega$ is
complex, and plane wave~(\ref{plane}) is unstable.

\begin{figure}[htb]
\centerline{
\includegraphics[width= 8cm]{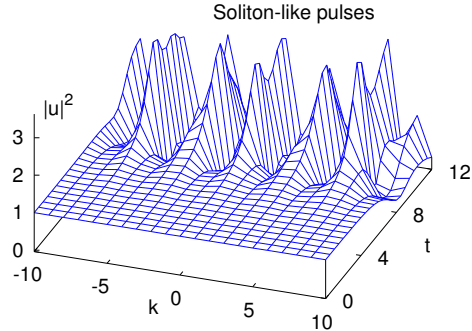}
}
\vspace{0.5cm}
\caption{The dynamics of a plane wave with random modulations.
The plane wave breaks up due to MI into a set of soliton-like pulses.}
\label{fig:mi}
\end{figure}

  Figure~\ref{fig:mi} shows the dynamics of a plane wave, modulated
initially by noise, $u_k(0) = 1 + \epsilon r_k$ , where $r_k$ is a random
number with uniform distribution in $[-1,1]$, and $\epsilon = 10^{-3}$.
Numerical simulations, as in Fig.~\ref{fig:mi}, can be obtained by commands,
similar to those in steps 1 and 4 (MC.1 and MC.4~\cite{Note}) of the
Mathematica code in~\ref{sec:app_code}, see also Problem 1 in Sec.~\ref{sec:prob}.

  Since noise has waves with arbitrary $Q$, some of modulation wavenumbers
are in the unstable region. As a result of instability, a plane wave
breaks into a set of soliton-like pulses. Though amplitudes of pulses
in Fig.~\ref{fig:mi} are varied, shapes and positions remain basically
the same. Therefore, MI indicates a presence of DSs in the system.
We analyze {\em stationary} DSs in the next Section.

\subsection{Discrete bright solitons}
\label{sec:ds}

  There is no exact DS solutions of the DNLS equation. One can construct
approximate solutions for different limiting cases. One limit is the case when
the soliton width is much larger than the distance between sites. In other
words, the variation of $u_n$ from one site to other is small. Then
Eq.~(\ref{dnls}) can be approximated by the continuous NLS equation:
\begin{equation}
  i {\partial w \over \partial t} + 2 \beta h^2
  {\partial^2 w \over \partial x^2} + \gamma |w|^2 w = 0,
\label{nls}
\end{equation}
where $w(x_k, t) \approx u_k(t) \exp(-2 \beta t)$, and $h$ is the distance
between neighbor sites. The NLS equation is the completely integrable
model~\cite{Ablo04} that has a vast number of exact solutions. We can use
the soliton solution of the NLS equation in order to construct an
approximate expression for the DS:
\begin{equation}
    u_{k,s}(t) = A\, \mbox{sech}(k / \nu )
    e^{i (2 \beta + \gamma A^2/2) t},
\label{solit}
\end{equation}
where  $\nu = (2 \beta)^{1/2} /(A \gamma^{1/2})$ is the soliton width, and
$A$ is a free parameter taken such that $\nu \gg 1$. In general, in the
analysis of a discrete system, it is useful to consider the properties of
its continuous counterpart.

   We mention that the continuous equation~(\ref{nls}) has bright soliton
solutions when $\beta \gamma > 0$~\cite{Kivs03,Ablo04}. In contrast, the
discrete counterpart~(\ref{dnls}) can have bright solitons for any sign of
$\beta \gamma$, provided that $d^2\omega/dq^2\, \gamma >0$. This is due to
the change of the dispersion sign, mentioned in Sec.~\ref{sec:pw}.

   Another limit for an approximate solution is the anti-continuum
limit~\cite{Kevr09}, when $\beta = 0$. In this case, the points are
decoupled from each other. Therefore, one can write a strongly localized
solution with only a single excited site
\begin{equation}
   u_k(t) = A\, \delta_{0, k}\, e^{i\gamma A^2 t}.
\label{ac_sol}
\end{equation}
When $\beta \neq 0$, this solution is not valid, but it can be used
as a basis solution in the perturbation theory or numerical simulations.

   An exact profile of a DS for an arbitrary set of the system parameters
can be found numerically. We look for the solution in the form $u_k(t) =
U_k \exp(-i \eta t)$. Then, the soliton profile $U_n$ and the corresponding
frequency $\eta$ are found from the nonlinear eigenvalue problem:
\begin{equation}
  - \beta (U_{k+1} + U_{k-1}) - \gamma |U_k|^2 U_k = \eta U_k ,
\label{nep}
\end{equation}
where $k = 1, \dots, N$. The derivation of this stationary equation can be
checked by using symbolic calculations, see MC.2.

  Eigenvalue problem~(\ref{nep}) has $N$ equations for $N+1$ unknowns,
namely, $\eta$ and $U_1, \dots, U_N$, subject to vanishing (zero or periodic)
boundary conditions for $U_k$. A direct numerical solution of such a
problem is a difficult task. However, if we fix the value of $\eta$,
then Eqs.~(\ref{nep}) are just a set of $N$ nonlinear algebraic equations
for $N$ unknowns. This set can easily be solved by corresponding routines,
see MC.3.

   For convergence of numerical solution of Eqs.~(\ref{nep}), proper $\eta$
and an initial distribution should be chosen. A value of $\eta$ should be
taken below the band of linear waves $\omega_{\mathrm{lin}} = [-2 \beta, 2
\beta]$. This can be deduced from the analysis of the dispersion
relation~(\ref{ldisp}), see also Problem 4 in Sec.~\ref{sec:prob}. An initial
distribution can be taken in a form close to a DS, for example, $U_k = A\,
\exp(-|k|)$, where $A$ is the soliton amplitude. An initial value of $A$ can
be taken as $A =(|\eta|/\gamma)^{1/2}$, cf. Eq.~(\ref{ac_sol}). However, we
find that the numerical procedure converges to the on-site (inter-site)
soliton, even when initially only one (two) site(s) is (are) excited.  Thus,
in the code, different solitons can be obtained by changing parameters
$\eta$, $c_1$ and $c_2$, see MC.1 and MC.3. Having a solution for one $\eta$,
one can restore the whole family of soliton solutions, by changing gradually
the value of $\eta$, see also Problem 6 in Sec.~\ref{sec:prob}.

  In order to solve the set of Eqs.~(\ref{nep}) for a given $\eta$, one can
use the Newton-Raphson method, see e.g.~\cite{Pres97}, which is the default
method of the Mathematica \texttt{\small FindRoot[]} command, see MC.3. One
can write Eqs.~(\ref{nep}) in a vector form $\mathbf{f}(\mathbf{z}) = 0$,
where $\mathbf{z} = (U_1, U_2, \dots, U_N)$, and $\mathbf{f}$ is  vector of
functions, $\mathbf{f} = (f_1, f_2, \dots, f_N)$. Then in the Newton-Raphson
method, a root is found by iterations~\cite{Pres97}
$\mathbf{z}_{\mathrm{new}} = \mathbf{z}_{\mathrm{old}} + \delta \mathbf{z}$,
where $\delta \mathbf{z}$ is a solution of a set of linear equations
$\mathbf{J}(\mathbf{z}_{\mathrm{old}})\, \delta \mathbf{z} =
-\mathbf{f}(\mathbf{z}_{\mathrm{old}})$, and $\mathbf{J}$ is the Jacobian
matrix $J_{nm} = \partial f_n / \partial U_m$.

   There are different types of discrete soliton
solutions~\cite{Kivs03,Flac08,Kevr09,Kart11,Lede01}. The distribution with
a center at a site is called the on-site soliton, while that with a center
between sites is called the inter-site soliton~\cite{Kivs03,Kevr09}. Both
on-site solitons and inter-site solitons can be symmetric or
anti-symmetric~\cite{Lede01}. If we define $u_k = (-1)^k v_k$, and change
$\gamma \to -\gamma$ and $t \to -t$, then dynamics of $v_k(t)$ is described
by the same DNLS equation. It means, in particular, that if $U_k$ is an
unstaggered mode with frequency $\eta < 0$ for $\gamma >0$, then $(-1)^k
U_k$ is a staggered mode with $\eta > 0$ for $\gamma <0$~\cite{Cai94}. This
is similar to plane wave profiles with $q= 0$ and $q = \pi$, respectively.
We consider only unstaggered symmetric solitons for $\gamma > 0$ (but see
also Problem 5 in Sec.~\ref{sec:prob}).

\begin{figure}[htb]
\centerline{
\includegraphics[width= 7.5cm]{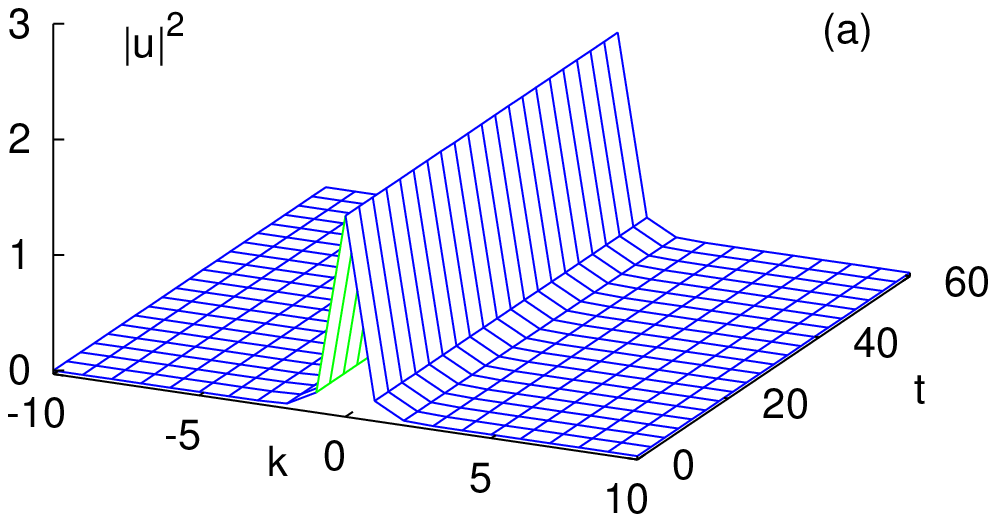} \hspace{4mm}
\includegraphics[width= 7.5cm]{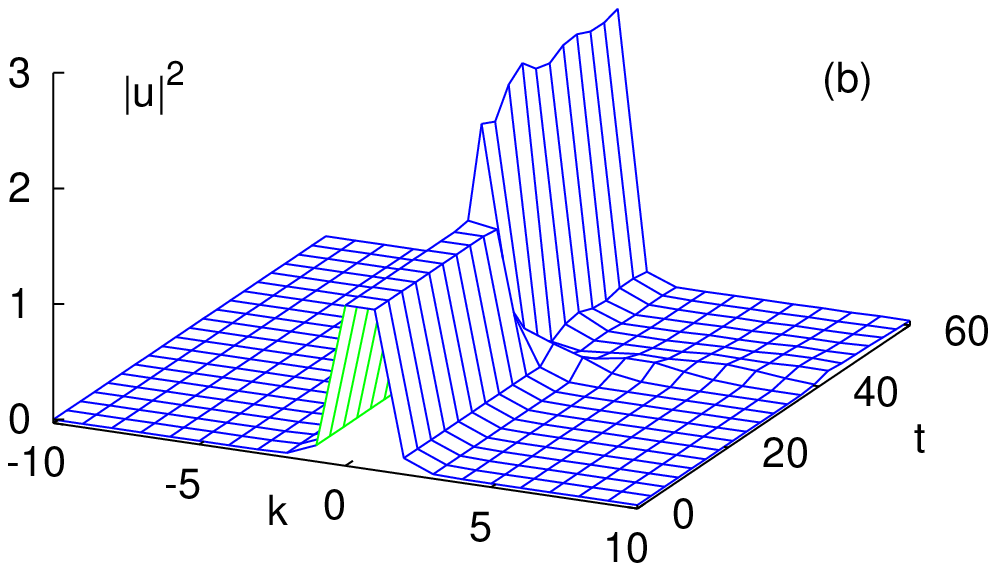}}
\caption{Dynamics of (a) on-site and (b) inter-site discrete solitons,
obtained from numerical simulations of Eq.~(\ref{dnls}). The system
parameters are $\beta= 0.5$, $\gamma = 1$, and $\eta= -2$.
}
\label{fig:sol}
\end{figure}

   We can check numerically that the solution found is indeed the
eigenmode. For this purpose, we integrate numerically Eq.~(\ref{dnls})
taking the solution of Eqs.~(\ref{nep}) as the initial condition, see MC.4.
We expect a stationary evolution at least for small $t$. Such numerical
simulations, using MC.4, show that on-site solitons are stable, while
inter-site are unstable~\cite{Kivs03,Kevr09}. Figure~\ref{fig:sol} shows
the dynamics of stable and unstable solitons.

\begin{figure}[htb]
\centerline{
\includegraphics[width= 7.cm]{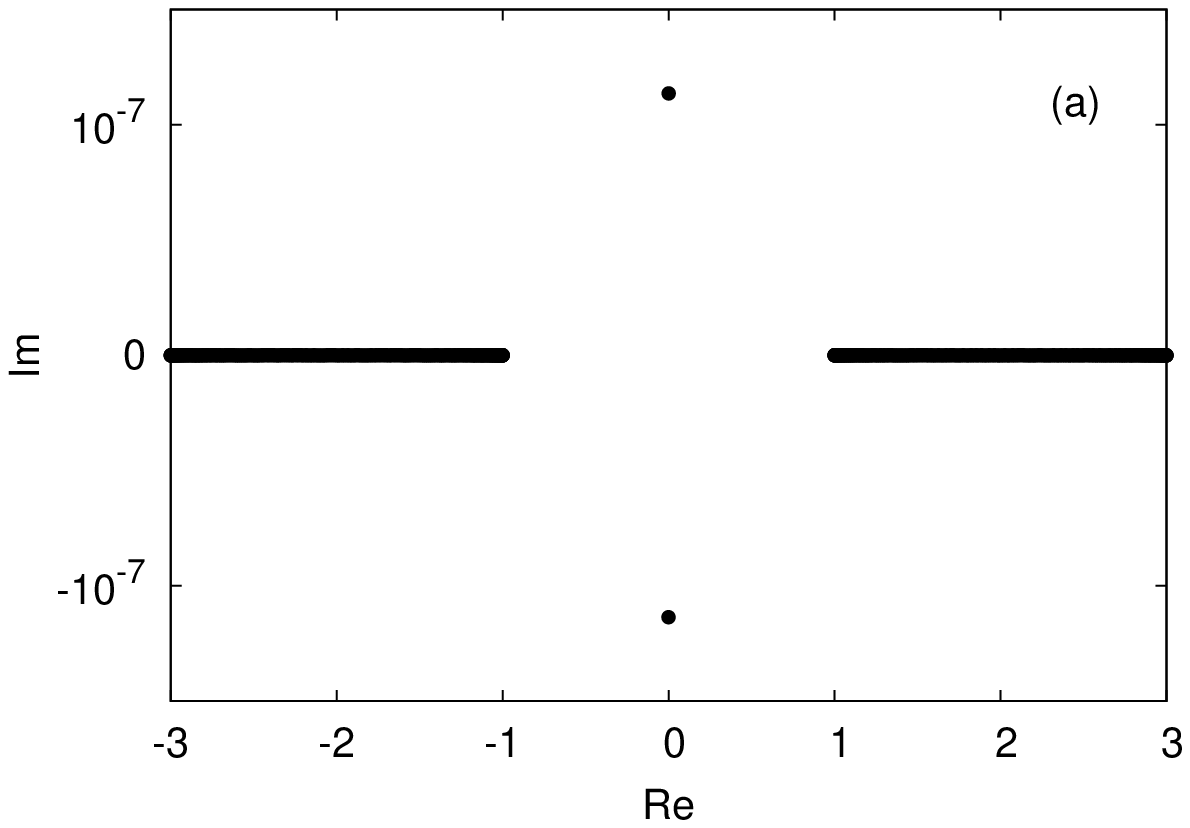}
\includegraphics[width= 7.cm]{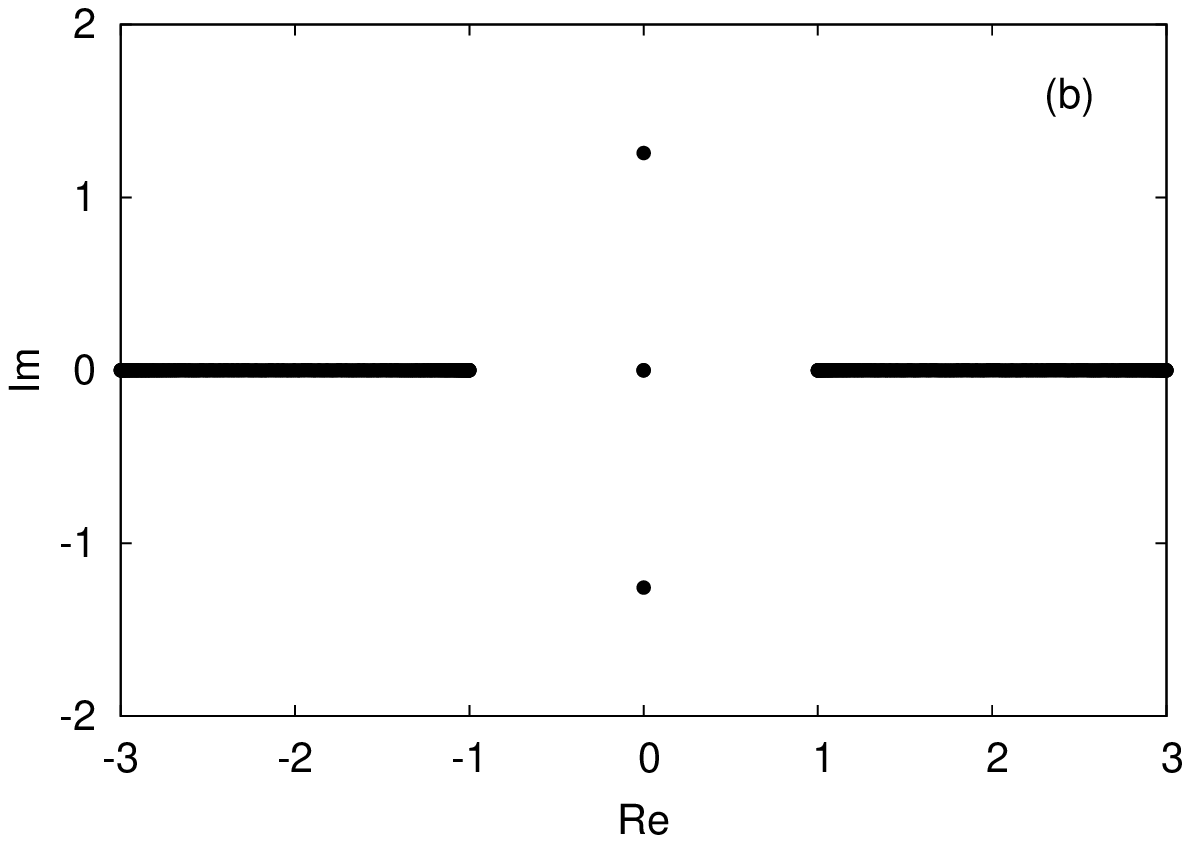}}
\caption{
  The distribution of modulation eigenvalues of the (a) on-site DS,
and (b) inter-site DS on the complex plane of $\Omega$.
The parameters are the same as in Fig.~\ref{fig:sol}.
}
\label{fig:evs}
\end{figure}

   Physically relevant localized waves correspond to stable solutions.
Therefore, it is necessary to analyze the stability of DSs. For this
purpose, similarly to Eq.~(\ref{pert}), we represent the solution as the
soliton with small perturbations $w_k(t)$ added:
\begin{equation}
  u_k(t) = [U_{k} + w_{k}(t)] \exp(-i \eta t).
\label{sol_pert}
\end{equation}
Then, in the first approximation, the evolution of perturbations is
described by the following equation:
\begin{equation}
   i {d w_k \over d t} + \eta w_k + \beta (w_{k+1} + w_{k-1}) +
     \gamma U_k^2 (2 w_k + w_k^{*}) = 0,
\label{eop}
\end{equation}
which can be checked with the code, see MC.5. Next, we apply a procedure
similar to that used for the analysis of plane waves. Namely, we obtain
from Eq.~(\ref{eop}) the equations for the real and imaginary parts of
$w_n$. Then substituting these parts in the form $\sim \exp(-i \Omega t)$,
we get a set of linear algebraic equations in a form $\mathbf{F\, A} = \Omega
\mathbf{A}$, where $\mathbf{A}$ is a real vector of length $2 N$ of
modulation amplitudes, and $\mathbf{F}$ is the matrix of coefficients.
Eigenvalues of $\mathbf{F}$ are the frequencies of modulations. A complex
frequency of modulations means instability of DSs. This stability analysis
is valid also for an infinite array, $N \to \infty$. All the steps
described are implemented in the code, see MC.6-MC.9. The distribution of
eigenvalues on the complex $\Omega$-plane are shown in Fig.~\ref{fig:evs}.
Figure~\ref{fig:evs}(a) shows that numerical accuracy of eigenvalue
calculations is of order $10^{-7}$.

   After identification of regions of stable DSs, one can perform a further
study of the discrete system. For example, one can study the evolution of
arbitrary localized distributions, or the interaction of discrete solitons.

\section{Questions and Problems}
\label{sec:prob}

  In this Section, we suggest few problems. The aim of the problems is to
help a deeper understanding of the methods reviewed in this paper, and also
to develop skills in working with Mathematica program in~\ref{sec:app_code}.
Problems 1-7 consider the fundamental properties of plane waves and solitons.
These properties are common for various systems. Solving these problems
provides more insight into the nonlinear dynamics of discrete systems.
Problems 8-12 and their extensions can be used as independent student
projects. The references at each problem are sources for further study.

  Problem 1~\cite{Kevr09,kivs1992}. Modify the program for the case of a
nonlinear plane wave to obtain a plot of modulational instability similar
to Fig.~\ref{fig:mi}.

  Problem 2~\cite{Kivs03,Kevr09}. The MI theory (see Eq.~(\ref{ds_pert}))
predicts an infinite growth of perturbations for some parameters. Does such
infinite growth occur in numerical simulations of the DNLS equation?

  Problem 3~\cite{Kivs03,Kevr09}. Prove analytically that $P$ and $H$ in
Eq.~(\ref{invar}) are indeed the invariants of Eq.~(\ref{dnls}). Include
to the program a calculation of $P$. This helps to monitor errors of
numerical simulations, see MC.4.

  Problem 4~\cite{Kivs03,Kevr09}. Analyze the dispersion relation~(\ref{ldisp}),
and figure out why the soliton frequency should satisfy $\eta < -2\beta$
($\beta > 0$). Hint: Consider the soliton center and tails as parts
of plane waves.

  Problem 5~\cite{Lede08,Lede01,Cai94}. Find anti-symmetric on-site and
inter-site solitons for $\gamma > 0$. Find staggered solions for $\gamma
<0$. Analyze their stability. For convergence, modify MC.3 to excite
initially 3-5 consecutive sites.

  Problem 6~\cite{Kivs03,Kevr09}. Using MC.1 and MC.3 of the program, find
the dependencies of the soliton amplitude and $P$ on $\eta$.

  Problem 7~\cite{Kivs93}. An initial condition in a form of
a localized mode (see Eq.~(\ref{nep}) and MC.3) multiplied by $\exp(i p k)$,
where $p$ is a constant, results in moving solitons. Modify step MC.4 of
the program to generate such solitons. How does the soliton velocity
depend on $\eta$ and $p$?

  Problem 8~\cite{Tsir94}. Consider a linear array with a nonlinear defect,
described by the following equation
\begin{equation}
  i {d u_k \over d t}  + \beta (u_{k+1}  + u_{k-1}) +
    \gamma \delta_{k,0} |u_k|^d u_k  = 0.
\label{defect}
\end{equation}
where $d$ is an arbitrary exponent. Find modes localized on the defect, and
check their dynamics and stability.

  Problem 9~\cite{Moli06}. Consider a semi-infinite array of waveguides,
described by Eq.~(\ref{dnls}) with $k > 0$ and $u_0 = 0$. Find and analyze
localized modes with maximum at different distances from the border ($k =
1$).

  Problem 10~\cite{carr2006,abdu2007}. Consider the cubic-quintic DNLS (C-Q
DNLS) equation, which can be represented in the following form
\begin{equation}
  i {d u_k \over d t}  + \beta (u_{k+1}  + u_{k-1}) +
    \gamma |u_k|^2 u_k+\delta|u_k|^4 u_k = 0.
\label{C-Q dnls}
\end{equation}
Modify and apply the program to find the different types of a localized
solutions of C-Q DNLS equation and analyze their linear stability
properties.

  Problem 11~\cite{Ablo04,salerno1,salerno2}. Consider the Ablowitz-Ladik
(A-L) equation
\begin{equation}
  i {d u_k \over d t}  + \beta (u_{k+1}  + u_{k-1}) +
    \gamma |u_k|^2(u_{k+1}  + u_{k-1}) = 0.
\label{A-L }
\end{equation}
This equation is an integrable discretization of the NLS equation and also
has some applications in physics. It has exact solitonic solutions. Modify
and apply the program to analyze the localized solutions of the A-L
equation. One can include the on-site cubic term to A-L equation (which
destroys the integrability) and analyze the properties of localized
solutions
\begin{equation}
  i {d u_k \over d t}  + \beta (u_{k+1}  + u_{k-1}) +
    \gamma |u_k|^2(u_{k+1}  + u_{k-1})+\delta|u_k|^2 u_k = 0.
\label{Salerno }
\end{equation}
This equation is called the Salerno equation.

  Problem 12~\cite{kov1}. The following inhomogeneous DNLS equation
describes the nonlinear localized impurity modes
\begin{equation}
  i {d u_k \over d t}  + \beta (u_{k+1}  + u_{k-1}) +
    \gamma |u_k|^2 u_k+\kappa \delta_{k,0}|u_k|^4 u_k = 0.
\label{C-Q dnls1}
\end{equation}
Use the Mathematica code to find these modes and check their stability.

\section{Conclusion}
\label{sec:concl}

   The basic steps in the study of discrete systems have been presented.
These steps include the construction of plane wave solutions and soliton
solutions of the discrete system, the derivation of equations for small
modulations, and analysis of stability. The corresponding code is
implemented in Mathematica. It is demonstrated that theoretical
consideration together with numerical modeling can substantially enhance
the understanding of properties of discrete systems. The approach described
here is quite generic, and it can be used to study other discrete systems.

   The DNLS equation considered does not include many effects. An actual
distributed system can be described by more general types of discrete
equations, some examples are considered in Sec.~\ref{sec:prob}. Further
extension of the DNLS model includes a consideration of two- and
three-dimensional arrays. Also, one can study higher-order interactions,
non-local effects, and different types of nonlinearity. One can take into
account external and parametric perturbations. We believe that the method
and the code presented in this work can serve as a starting point to study
these systems.

\ack

   The work by E. N. T. has been supported by grant FA-F2-004 of the
Agency for Science and Technologies, Uzbekistan,
and the work by B. A. U. has been partially supported by project FRGS
16-014-0513 of the  Ministry of Higher Education, Malaysia.

\appendix
\section{Mathematica code}
\label{sec:app_code}

   Here we describe a code in Mathematica (ver.8.0) for symbolic and
numerical analysis of Eq.~(\ref{dnls}). A Supplementary File (both in np- and
pdf-formats) provides the code with outputs, so that it gives a working
example for a particular set of parameters. The code is valid for equations
with nearest neighbor terms, however it can easily be extended to more
general cases. Steps MC.1-MC.4 show a finding of a localized mode and a
numerical simulation of its dynamics, while MC.5-MC.10 are related to a
derivation of equations for modulations and calculation of the modulation
spectrum.

\vspace{6pt}
1. Define parameters and the DNLS equation (eqn):

\noindent \texttt{\small
nPoints=200;  tend=50;  \\
par1= \{$\beta$-> 0.5, $\gamma$-> 1, $\omega$-> -2\};  par2= \{c1-> 1., c2-> 1.\}; \\
eqn= I*u[k]'[t] + $\beta$*(u[k+1][t]+u[k-1][t]) + $\gamma$*(u[k][t])$^\wedge$2*Conjugate[u[k][t]]
}

\vspace{6pt}
2. Derive the stationary equation (eqnStat) from the DNLS equation :

\noindent \texttt{\small
sub1= u[k\_ ]-> Function[\{t\}, U[k]*Exp[-I*$\omega$*t]]; \\
eqn1= Simplify[eqn /.sub1, Assumptions-> \{\{$\omega$, t, U[i\_ ]\} $\in$ Reals\}]; \\
eqnStat= Coefficient[eqn1, Exp[-I*$\omega$*t]]
}

\vspace{6pt}
3. Find eigenmode (eigenMode) from the list of stationary equations (eqnStatList):

\noindent \texttt{\small
eqnStatList= Table[eqnStat, \{k,1,nPoints\}] /.  \\
\phantom{  }  \{U[0]-> U[nPoints], U[nPoints+1]-> U[1]\};  \\
init= Table[\{U[k], (c1*KroneckerDelta[nPoints/2,k] +  \\
\phantom{  }    c2*KroneckerDelta[nPoints/2+1,k]) /.par2\}, \{k,1,nPoints\}];  \\
eigenMode= FindRoot[(eqnStatList /.par1)==0, init];  \\
init1= Table[U[k], \{k,1,nPoints\}] /.eigenMode;  \\
ListLinePlot[init1, PlotRange-> \{\{nPoints/2-10, nPoints/2+10\}, All\},  \\
\phantom{  }   PlotMarkers-> Automatic]
}

\vspace{6pt}
4. Solve the DNLSE, using eigenmode as an initial condition:

\noindent \texttt{\small
initialCond= Table[u[k][0]==init1[[k]], \{k,1,nPoints\}];  \\
dnlsList= Join[Table[(eqn /.par1)==0, \{k,1,nPoints\}] /.  \\
\phantom{  }  \{u[0][t]-> u[nPoints][t], u[nPoints+1][t]-> u[1][t]\}, initialCond];  \\
sol1= NDSolve[dnlsList, Table[u[k], \{k,1,nPoints\}], \\
\phantom{  }  \{t, 0, tend\}, MaxSteps-> 1000000]; \\
fig1= Evaluate[Table[Abs[u[k][t]], \{t, 0, tend\}, \{k,1,nPoints\}] /.sol1]; \\
ListPlot3D[fig1, PlotRange-> All]
}

\vspace{6pt}
5. Derive an equation for the first correction, w[i][t] :

\noindent \texttt{\small
sub1= u[k\_ ]-> Function[\{t\}, (U[k] + $\epsilon$*w[k][t]) Exp[-I*$\omega$*t]]; \\
res1= Coefficient[Simplify[eqn /.sub1, Assumptions-> \\
\phantom{  }  \{\{$\omega$, t, U[i\_ ], $\epsilon$\} $\in$ Reals, w[i\_ ][t]
  $\in$ Complexes\}], Exp[-I*$\omega$*t]]; \\
eqnFirstCorr= Coefficient[Collect[res1, $\epsilon$], $\epsilon$, 1]
}

\vspace{6pt}
6. Split equation eqnFirstCorr  into real and imaginary parts (wr, wi):

\noindent \texttt{\small
res1= Simplify[eqnFirstCorr /.w[k\_ ] ->  Function[\{t\}, (wr[k][t]+I*wi[k][t])], \\
\phantom{  }  Assumptions-> \{wr[k\_ ][t] $\in$ Reals, wi[k\_ ][t] $\in$ Reals\}]; \\
eqnz1r= Simplify[ComplexExpand[Im[res1]]] \\
eqnz1i= Simplify[ComplexExpand[Re[res1]]]
}

\vspace{6pt}
7. Find left-hand sides (lhs1r and lhs1i) of the equations for spatial distributions:

\noindent \texttt{\small
sub2= \{wr[k\_ ]-> Function[\{t\}, ar[k]*Exp[-I*$\Omega$*t]],  \\
\phantom{  }  wi[k\_ ]-> Function[\{t\}, ai[k]*Exp[-I*$\Omega$*t]]\};  \\
eqn1r= Expand[Coefficient[Simplify[eqnz1r /.sub2], Exp[-I*$\Omega$*t], 1]];  \\
eqn1i= Expand[Coefficient[Simplify[eqnz1i /.sub2], Exp[-I*$\Omega$*t], 1]];  \\
lhs1r= Simplify[$\Omega$*ar[k] - eqn1r/Coefficient[eqn1r, $\Omega$*ar[k], 1]]  \\
lhs1i= Simplify[$\Omega$*ai[k] - eqn1i/Coefficient[eqn1i, $\Omega$*ai[k], 1]]
}

\vspace{6pt}
8. Construct a matrix of coefficients in a difference form (matrF) and a vector
of unknowns (vectA):

\noindent \texttt{\small
sub3= \{ai[0]-> ai[nPoints], ai[nPoints+1]-> ai[1],  \\
\phantom{  }  ar[0]-> ar[nPoints], ar[nPoints+1]-> ar[1]\};  \\
matrFdiff= Join[Table[lhs1r,\{k,1,nPoints\}] /.sub3,  \\
\phantom{  } Table[lhs1i, \{k,1,nPoints\}] /.sub3]; \\
vectA= Join[Table[ar[k], \{k,1,nPoints\}], Table[ai[k], \{k,1,nPoints\}]];
}

\vspace{6pt}
9. Substitute eigenmode and parameters into matrFdiff, and convert it to
a {``}standard{''} form. Then find eigenvalues of matrix F:

\noindent \texttt{\small
m1= matrFdiff /.eigenMode /.par1;  \\
\{bb1, matrF\}= CoefficientArrays[m1, vectA]; \\
ev1= Eigenvalues[Normal[matrF]];  Max[Im[ev1]]
}

\vspace{6pt}
10. Plot eigenvalues on the complex plane:

\noindent \texttt{\small
ev1Fig= Table[\{Re[ev1[[i]]], Im[ev1[[i]]]\}, \{i,1,2*nPoints\}];  \\
ListPlot[ev1Fig, PlotRange-> All, PlotMarkers->  \\
\phantom{  } Graphics[\{Blue,Thick,Circle[]\}, ImageSize-> 8]]
}


\begin{thebibliography}{99}
\bibitem{Note} In the text, for brevity we use notation ``MC.$n$'' that
refers to $n$-th step of the Mathematica code in Appendix.
\end{thebibliography}
\end{document}